\documentclass[aps,prd,amsmath,amssymb,preprintnumbers,onecolumn,11pt,nofootinbib]{revtex4}
\usepackage[a4paper,left=20mm,right=20mm,top=25mm,bottom=25mm]{geometry}
\usepackage{amsmath,amssymb}
\usepackage{mathrsfs}
\usepackage{graphicx}
\usepackage{color}
\usepackage{subfigure}
\usepackage{pxfonts}
\usepackage{fancyhdr}
\usepackage{multirow}
\usepackage{mathpazo}

\definecolor{rossoCP3}{cmyk}{0,0.88,0.77,0.40}
\newcommand{\symb}{{\color{rossoCP3}{\vardiamondsuit}}}

\newcommand{\tr}[1]{\langle #1\rangle}
\newcommand{\cL}{\mathcal{L}}
\newcommand{\cO}{\mathcal{O}}
\newcommand{\Jbar}{\overline{J}}
\newcommand{\Hbar}{\overline{H}}

\begin{document}

\title{\Large \color{rossoCP3} Extending Chiral Perturbation Theory with an Isosinglet Scalar}
\author{Martin Hansen$^\symb$}
\email{hansen@cp3.sdu.dk}
\author{Kasper Lang\ae ble$^\symb$}
\email{langaeble@cp3.sdu.dk} 
\author{Francesco Sannino$^\symb$}
\email{sannino@cp3.sdu.dk} 
\affiliation{\footnotesize $^\symb${ CP}$^3${-Origins}, University of Southern Denmark, Campusvej 55, DK-5230 Odense M, Denmark.}

\begin{abstract}
We augment the chiral Lagrangian by an isosinglet scalar and compute the one-loop radiative corrections to the pion mass and decay constant, as well as the scalar mass. The calculations are carried out for different patterns of chiral symmetry breaking of immediate relevance for phenomenology and lattice investigations. By construction our results encompass several interesting limits, ranging from the dilaton to the linear sigma model. \\[2mm]
{\footnotesize\it Preprint: CP$^3$-Origins-2016-041 DNRF90}
\end{abstract}

\maketitle

\section{Introduction}
There has been a growing interest in realizing and studying possible strongly coupled theories featuring an isosinglet scalar as the first massive state appearing after the (pseudo) Goldstone bosons. At energy scales much below the isosinglet mass, chiral perturbation theory holds true for massless Goldstone bosons, and the scalar field can safely be integrated out. However, if one is interested in the dynamics involving energy scales near or around the isosinglet state, or if its mass is close to the pion mass, its quantum effects cannot be neglected, and the isosinglet state must be integrated back in.

For this reason we consider the chiral Lagrangian augmented by an isosinglet scalar and show how this leads to new radiative corrections for the pion mass, the pion decay constant, and the scalar mass. As we will argue, these corrections depend on the number of Goldstone bosons, but are otherwise universal in form for all patterns of chiral symmetry breaking. We focus on two patterns of chiral symmetry breaking i.e. $SU(2)_L\times SU(2)_R\to SU(2)_V$ and $SU(4)\to Sp(4)$. The first breaking pattern has direct relevance for the interpretation of the $f_0(500)$ in QCD  \cite{Sannino:1995ik,Harada:1995dc,Harada:1996wr,Harada:2003em,Black:2000qq,Caprini:2005zr,GarciaMartin:2011jx,Parganlija:2012fy,Pelaez:2015zoa,Ghalenovi:2015mha} as well as the emergence of a potentially light scalar state in near conformal theories with two Dirac fermions in a complex representation of the gauge group, such as the two-index symmetric representation of $SU(3)$ \cite{Sannino:2004qp,Dietrich:2005jn,Sannino:2009za}. The second breaking pattern emerges when two Dirac fermions belong to the fundamental representation of $Sp(2N)$ which for $N=1$ corresponds to $SU(2)$. This theory became the ideal template for numerous relevant extensions of the standard model, ranging from ultraminimal technicolor \cite{Appelquist:1999dq,Dietrich:2006cm,Ryttov:2008xe} to composite (Goldstone) Higgs \cite{Cacciapaglia:2014uja,Arbey:2015exa}, as well as strongly interacting massive particles (SIMPs) for dark matter \cite{Hochberg:2014kqa,Hansen:2015yaa}.

Lattice simulations are currently investigating these models \cite{Briceno:2016mjc,Fodor:2012ty,Fodor:2015zna,Hasenfratz:2015ssa,Arthur:2016dir,Arthur:2016ozw} and they can therefore directly compare their results with our findings once the spectrum is known precisely enough. It is furthermore straightforward to generalise our results to the $SU(N_f) \times SU(N_f) \rightarrow SU(N_f)$ chiral symmetry breaking pattern. 

To organise perturbation theory we adopt the power counting scheme $\cO(p)\sim\cO(m_\pi)\sim\cO(m_\sigma)\ll \Lambda_\chi$ where $\Lambda_\chi$ is the scale of chiral symmetry breaking, expected to be of the order $4\pi f_\pi$. The chosen counting scheme is tailored to properly account for a light scalar state, henceforth limiting the applicability for heavier scalar states. According to this scheme the leading order (LO) corresponds to $\cO(p^2)$ and the next-to-leading order (NLO) corresponds to $\cO(p^4)$. Previous investigations have already appeared in the literature, e.g. \cite{Soto:2011ap}. We will generalise this analysis by extending the set of operators present at the tree-level Lagrangian and by considering different patterns of chiral symmetry breaking. 

Having introduced a holistic approach for the scalar field, we then consider different realisations, such as the dilaton, the Goldstone boson, and linear sigma model.

The paper is structured as follows: In Section II we introduce the Lagrangian and the renormalisation procedure used to subtract divergences. In Section III we present the one-loop corrections to the pion mass, the pion decay constant, and the scalar mass. We also perform several consistency checks to ensure that we can reproduce known results in different limits. In Section IV we consider different realisations of the scalar field and show how this leads to constraints on the low-energy constants.

\section{The Lagrangian}
\label{sec:lagrangian}
In this section we introduce the non-linearly realized chiral Lagrangian augmented with an isosinglet scalar. We follow the notation of \cite{Bijnens:2009qm} and let $G$ be the global flavor symmetry of the vector-like fermions and $H$ the stability group after spontaneous symmetry breaking. The Goldstone boson manifold $G/H$ is then parametrized by
\begin{equation}
 u = \exp\left(\frac{i}{\sqrt{2}f_\pi}X^a\phi^a\right),
\end{equation}
where $X^a$ are the broken generators. In our convention the generators are normalized as $\tr{X^aX^b}=\delta^{ab}$ where $\tr{\cdot}$ denotes trace in flavor space. The quantity $u$ transforms as
\begin{equation}
 u \to g u h^\dagger = h u g^\dagger,
\end{equation}
with $g\in G$ being space-time independent and $h\in H$ being space-time dependent in such a way that the above constraint equation is satisfied. The linear realization \cite{Gasser:1983yg,Gasser:1984gg} of the chiral Lagrangian is parametrized in terms of the field $U=u^2$ which transforms under the global symmetry $G$ instead of the stability group $H$. The quantities that transform homogeneously under the stability group $H$ are
\begin{align}
 u_\mu &= i(u^\dagger(\partial_\mu-ir_\mu)u - u(\partial_\mu-il_\mu)u^\dagger), \\
 \chi_\pm &= u^\dagger\chi u^\dagger \pm u\chi^\dagger u.
\end{align}
In the first expression $r_\mu$ and $l_\mu$ are external currents, which are needed e.g. when calculating the corrections to the pion decay constant. In the second expression $\chi$ is a spurion field that ensures chiral invariance at every step of the computation. The precise definition of $r_\mu, l_\mu$ and $\chi$ as a function of the external fields are given for each of the breaking patterns in \cite{Bijnens:2009qm}. In the end, 
the field $\chi$ is replaced by its expectation value $\chi=m_\pi^2$ which explicitly breaks the chiral symmetry. In the isospin limit, the leading-order pion mass can be written as $m_\pi^2=2B_0m_q$ where $B_0$ is related to the underlying chiral condensate and $m_q$ is the quark mass. In this notation the LO Lagrangian is given by
\begin{equation}
 \cL_2 = \frac{f_\pi^2}{4}\tr{u_\mu u^\mu + \tilde{\chi}_+} , 
\end{equation}
where $\tilde{\chi}_+ = \chi_+ - (\chi + \chi^\dagger)$. In the definition of $\tilde{\chi}_+$ we subtract a constant term to avoid mixing between the vacuum and the scalar field later on. The NLO Lagrangian reads 
\begin{align}
\begin{split}
 \cL_4
 &= L_0\tr{u_\mu u_\nu u^\mu u^\nu}
 + L_1\tr{u_\mu u^\mu}\tr{u_\nu u^\nu}
 + L_2\tr{u_\mu u_\nu}\tr{u^\mu u^\nu}
 + L_3\tr{u_\mu u^\mu u_\nu u^\nu} \\
 &\quad + L_4\tr{u^\mu u_\mu}\tr{\chi_+}
 + L_5\tr{u^\mu u_\mu\chi_+}
 + L_6\tr{\chi_+}^2
 + L_7\tr{\chi_-}^2
 + \tfrac{1}{2}L_8\tr{\chi_+^2+\chi_-^2}.
\end{split}
\end{align}
This parametrization differs from the one of \cite{Gasser:1983yg,Gasser:1984gg} and for this reason the low-energy constants (LECs) cannot be directly compared. However, they can be related though a careful mapping between the two parametrizations. Depending on the specific pattern of chiral symmetry breaking, some of the operators in the Lagrangian can become linearly dependent, and this is the case for the two specific patterns studied here. We choose not to reduce the number of operators in the Lagrangian because our results can be applied to a wider class of theories. We finally note that the $L_7$ term does not contribute at NLO in the isospin preserving limit. Because the NLO Lagrangian represents the most general Lagrangian at $\cO(p^4)$ it is possible to absorb the one-loop divergences by an appropriate renormalization of the LECs. We use the modified minimal-subtraction scheme ($\overline{\mathrm{MS}}$) where
\begin{equation}
 L_i = L_i^r - \frac{\Gamma_i}{32\pi^2}R,
 \label{eq:Lr}
\end{equation}
with
\begin{equation}
 R = \frac{2}{\epsilon} + \log(4\pi) - \gamma_E + 1.
\end{equation}
Here $\epsilon=4-d$ and $\gamma_E=-\Gamma'(1)$ is the Euler-Mascheroni constant. It should be noted that the renormalized coefficients $L_i^r$ will depend on the energy scale $\mu$ introduced by dimensional regularization.

We will now introduce the isosinglet scalar $\sigma$ in the chiral Lagrangian as a non-trivial background field \cite{Cacciapaglia:2014uja,Sannino:2015yxa}. In practice this is done by expanding each coefficient in the Lagrangian in powers of $\sigma/f_\pi$. Because we are interested in calculating the radiative corrections to the two-point functions at next-to-leading order, the expansion is only needed for the leading-order Lagrangian and we can stop the series expansion at second order.
\begin{equation}
 \cL_2
 = \frac{f_\pi^2}{4}\left[1+S_1\left(\frac{\sigma}{f_\pi}\right)+S_2\left(\frac{\sigma}{f_\pi}\right)^2\right]\tr{u_\mu u^\mu} 
 + \frac{f_\pi^2}{4}\left[1+S_3\left(\frac{\sigma}{f_\pi}\right)+S_4\left(\frac{\sigma}{f_\pi}\right)^2\right]\tr{\tilde{\chi}_+}
\end{equation}
The associated Lagrangian for the scalar field can be written as
\begin{equation}\label{scalarL}
 \cL_\sigma
 = \frac{1}{2}\partial_\mu\sigma\partial^\mu\sigma
 - \frac{1}{2}m_\sigma^2\sigma^2\left[1+S_5\left(\frac{\sigma}{f_\pi}\right)+S_6\left(\frac{\sigma}{f_\pi}\right)^2\right].
\end{equation}
In the Lagrangian for the scalar field we could perform a similar expansion in front of the kinetic term. However, in the present analysis these additional terms would correspond to a shift of  $S_5$ and $S_6$  because we only consider on-shell quantities. In our approach we assume that the scalar field has vanishing expectation value, $\tr{\sigma}=0$, and this leads to certain constraints on the two parameters $S_5$ and $S_6$ controlling the potential.
\begin{equation}
 S_5 \geq-2\sqrt{S_6},\qquad S_6\geq0.
\end{equation}

We are now ready to move to the renormalization procedure which follows the standard route of quantum correcting the theory when enforcing the counting $\cO(p)\sim\cO(m_\pi)\sim\cO(m_\sigma)$. To cancel the one-loop divergences we need to introduce a set of counter terms. Aside from using the NLO low-energy constants $L_i$ to cancel divergences, when introducing the scalar it will be necessary to introduce additional counter terms. In the equation below, the first two terms correspond to renormalizing $f_\pi$ and $B_0$ and they are needed to cancel divergences in the pion mass and the pion decay constant. The remaining counter terms are needed to cancel divergent contributions to the scalar mass.
\begin{equation}
 \cL_{CT}
 = K_1m_\sigma^2\tr{u_\mu u^\mu}
 + K_2m_\sigma^2\tr{\chi_+}
 + \frac{1}{f_\pi^2}\left(K_3(\partial^2\sigma)^2
 + K_4m_\pi^2(\partial_\mu\sigma)^2
 + K_5m_\pi^4\sigma^2\right) \ .
 \label{eq:Lct}
\end{equation}
For convenience we write the counter terms with appropriate factors of either the scalar or pion mass, because it allows us to keep the convention used in Eq.~\eqref{eq:Lr}.
\begin{equation}
 K_i = K_i^r - \frac{\Gamma^K_i}{32\pi^2}R \ .
 \label{eq:Kr}
\end{equation}
Setting the finite part $K_i^r$ to zero is allowed because it corresponds to a redefinition of the remaining coefficients and bare quantities in the Lagrangian. However, we keep them as unspecified constants in the calculations because they are needed when discussing renormalization scale dependence.

As stated in the introduction, the structure of the scalar contributions to the pion mass and decay constant have a universal structure at next-to-leading order. The origin of this universality resides in the fact that, at the lowest relevant order, the interactions of the scalar field involve either the pion kinetic term or the pion mass term as shown below.
\begin{equation}\label{expandedL}
 \cL_2
 = \frac{1}{2}\left[1+S_1\left(\frac{\sigma}{f_\pi}\right)+S_2\left(\frac{\sigma}{f_\pi}\right)^2\right](\partial_\mu\phi\cdot\partial^\mu\phi)
 - \frac{1}{2}\left[1+S_3\left(\frac{\sigma}{f_\pi}\right)+S_4\left(\frac{\sigma}{f_\pi}\right)^2\right]m_\pi^2(\phi\cdot\phi)
\end{equation}
For the pion decay constant one can use the operators associated to the external left transforming current\footnote{In the pseudo-real case, the external field coupling to left-handed quarks enter in both $r_\mu$ and $l_\mu$ at the effective level. However, since the interactions with the isosinglet scalar factorize, this distinction will not make a difference at NLO.}, which again is universal at lowest order.
\begin{equation}
 \cL_2 = \frac{f_\pi}{\sqrt{2}}\left[1+S_1\left(\frac{\sigma}{f_\pi}\right)+S_2\left(\frac{\sigma}{f_\pi}\right)^2\right](\partial_\mu\phi\cdot l^\mu) \ .
\end{equation}
Let us now pause and summarise  the three sets of low-energy constants present in the outlined set-up. The first set $L_i^r$ parametrises the pion interactions in the original chiral Lagrangian, and their values are known in QCD \cite{Bijnens:2014lea}. The LECs can in general be divided in contributions from the heavier resonances $R$ that have been integrated out in the effective theory, plus a remaining piece. 
\begin{equation}\label{LRes}
L_i^r = \hat{L}_i + \sum_{R}\,L_i^R \ ,
\end{equation}
In QCD \cite{Gasser:1983yg,Ecker:1988te,Donoghue:1988ed} it was argued that   heavy spin one resonances saturate the right-hand side of Eq.~\eqref{LRes}, meaning that the remainder $\hat L_i$ is subleading compared to $L_i^r$, for certain processes. In these papers the lightest QCD scalar resonance was assumed to be heavy and therefore integrated out. In our approach the scalar resonance is assumed to be light and for this reason it cannot be integrated out. As a result, the sum in Eq.~\eqref{LRes} will no longer include the isosinglet scalar contribution, and the theory will furthermore be affected by the presence loops with a light propagating scalar. For this reason the $L_i^r$ coefficients will in general be different. When increasing the mass of the scalar resonance, the present framework recovers the results of \cite{Gasser:1983yg,Ecker:1988te,Donoghue:1988ed} in terms of $S_1$ and $S_3$. Operators with higher powers of the scalar field, explicitly the operators proportional to the couplings $S_2$ and $S_4$, are subleading in this limit.

The second set $S_i$ parametrises the scalar interactions, and the last set $K_i^r$ is associated to the counter terms needed to cancel divergences. From naive dimensional analysis, it is natural to expect that NLO operators should be suppressed by a loop factor $1/(4\pi)^2\sim10^{-3}$ and this allows us to estimate the size of the low-energy constants.
\begin{equation}
 L_i^r \sim \cO(10^{-3}), \qquad S_i \sim \cO(1), \qquad K_i^r \sim \cO(10^{-3}).
 \label{eq:LKSest}
\end{equation}
For QCD, this naive estimate for $L_i^r$ is in agreement with the results from lattice simulations and experiments. For this reason, we believe that similar estimates should hold true for $S_i$ and $K_i^r$.

\section{NLO Corrections}
\subsection{Pion Mass and Decay Constant}
\begin{figure}
 \begin{center}
  \includegraphics[width=0.20\linewidth]{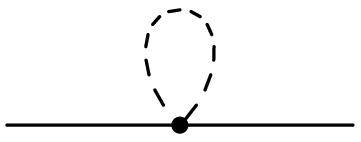}
  \hspace{2mm}
  \includegraphics[width=0.20\linewidth]{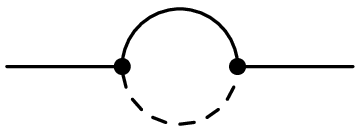}
  \hspace{7mm}
  \includegraphics[width=0.20\linewidth]{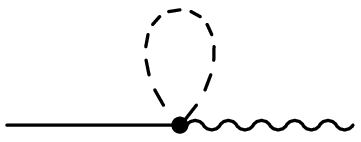}
  \hspace{2mm}
  \includegraphics[width=0.20\linewidth]{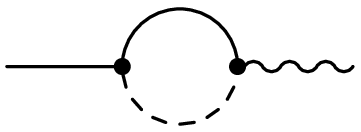}
  \caption{Loop diagrams contributing (at next-to-leading order) to the pion mass (left) and pion decay constant (right). The solid lines are pions, the dashed lines are scalars, and the wiggly lines are external currents. Here we only show the new diagrams involving the scalar field, but in both cases there is an additional tadpole diagram and a contact term.}
  \label{fig:pion_diagrams}
 \end{center}
\end{figure}

Since the physical pion mass is defined as the pole in the propagator, we now  determine the contributions to the pion self-energy and solve for the pole-mass via the equation
\begin{equation}
 \hat{m}_\pi^2 - m_\pi^2 - \Sigma(\hat{m}_\pi^2) = 0,
\end{equation}
where $\hat{m}_\pi^2$ is the physical pion mass, $m_\pi^2$ is the bare pion mass in the Lagrangian, and $\Sigma(p^2)$ is the pion self-energy. When including the scalar field there are two new diagrams, as shown in Fig.~\ref{fig:pion_diagrams} (left), contributing to the self-energy. Similarly there are two new diagrams contributing to the pion decay constant, as shown in Fig.~\ref{fig:pion_diagrams} (right), where the outgoing legs have been replaced by an external current.

Before we write down the results for $\hat{m}_\pi^2$ and $\hat{f}_\pi$ we introduce some short-hand notation to make  the results more readable. We write the chiral logs as
\begin{equation}
 L_x = \frac{1}{16\pi^2}\log\left(\frac{m_x^2}{\mu^2}\right),
\end{equation}
with $x=\{\pi,\sigma\}$ denoting one of the two masses. From the loop diagrams we obtain unitarity corrections written in terms of the functions $\Jbar(m_1^2,m_2^2,p^2)$ and $\Hbar(m_1^2,m_2^2,p^2)$ defined in Appendix~\ref{sec:loop_integrals}. For the on-shell results we use the auxiliary functions
\begin{equation}
 J_{xyz} = \frac{1}{16\pi^2}\Bigg[\Jbar(m_x^2,m_y^2,m_z^2)+1\Bigg],\qquad
 H_{xyz} = \frac{1}{16\pi^2}\Bigg[\Hbar(m_x^2,m_y^2,m_z^2)\Bigg],
\end{equation}
to shorten the expressions.\\
In this notation, the pion mass now reads
\begin{align}
\begin{split}
 \hat{m}_\pi^2
 = m_\pi^2
 + \frac{m_\pi^4}{f_\pi^2}(a_1+a_2L_\pi+a_3J_{\pi\sigma\pi})
 &+ \frac{m_\sigma^4}{f_\pi^2}(a_4L_\sigma + a_5J_{\pi\sigma\pi}) \\
 &\quad+ \frac{m_\pi^2m_\sigma^2}{f_\pi^2}(a_6 + a_7L_\pi+a_8L_\sigma+a_9J_{\pi\sigma\pi}),
 \label{eq:Mhat}
\end{split}
\end{align}
and the pion decay constant is
\begin{align}
\begin{split}
 \hat{f}_\pi
 = f_\pi + \frac{m_\pi^2}{f_\pi}(b_1 + b_2L_\pi + b_3J_{\pi\sigma\pi})
 &+ \frac{m_\sigma^2}{f_\pi}(b_4 + b_5L_\sigma + b_6J_{\pi\sigma\pi}) \\
 &\quad+ \frac{H_{\pi\sigma\pi}}{f_\pi}(b_7m_\pi^4 + b_8m_\sigma^4 + b_9m_\pi^2m_\sigma^2).
 \label{eq:Fhat}
\end{split}
\end{align}
From the above it is evident that the presence of the scalar field dramatically increases the complexity of the resulting  corrections when comparing to the usual ChPT case, where the only non-zero coefficients are $a_{1,2}$ and $b_{1,2}$. The coefficients $a_i$ and $b_i$ are combinations of the low-energy constants $S_i$ as shown in Table~\ref{table:abc}. In the definition of $a_{1,2}$ and $b_{1,2}$ the coefficients $a_{M,F}$ and $b_{M,F}$ encode the standard results from ChPT, which depend on the symmetry breaking pattern. For the two symmetry breaking patterns considered here, these coefficients can be found in Table~\ref{table:gamma} together with the values of $\Gamma_i$ and $\Gamma_i^K$ needed to cancel the divergences. In the general case, they can be found in \cite{Bijnens:2009qm}.

For the pion decay constant the last term is special and it arises because we need to take the derivative of $\Jbar(m_1^2,m_2^2,p^2)$ when calculating the renormalization constant. One should note that $H_{\pi\sigma\pi}$ has mass dimension minus two and this is the reason for the additional powers of mass multiplying this term.

\begin{table}
\begin{tabular}{c|c|c|c|}
 $\qquad$
 & $\hspace{18mm}a_i\hspace{18mm}$
 & $\hspace{15mm}b_i\hspace{15mm}$
 & $\hspace{22mm}c_i\hspace{22mm}$ \\
 \hline
 1 & $b_M$
   & $b_F$
   & $6S_6$ \\
 2 & $a_M-\tfrac{1}{2}(2S_1S_3-S_1^2)$
   & $a_F-\tfrac{3}{8}S_1^2$
   & $-\tfrac{1}{4}n_\pi S_1^2$ \\
 3 & $-(S_1-S_3)^2$
   & $-\tfrac{1}{2}(S_1S_3-S_1^2)$
   & $-9S_5^2$ \\
 4 & $\tfrac{1}{4}S_1^2$
   & $-2K_1^r$
   & $2n_\pi(S_1S_3-S_1^2) + n_\pi(S_4-S_2)$ \\
 5 & $-a_4$
   & $\tfrac{1}{8}(12S_2+S_1^2)$
   & $-n_\pi(S_1-S_3)^2$ \\
 6 & $4(K_2^r-K_1^r)$
   & $-\tfrac{1}{4}S_1^2$
   & $\tfrac{1}{2}n_\pi S_1^2$ \\
 7 & $-a_4$
   & $-\tfrac{1}{2}(S_1-S_3)^2$
   & $n_\pi(S_1^2-S_1S_3)$ \\
 8 & $S_4-S_2-S_1^2+S_1S_3$
   & $-\tfrac{1}{8}S_1^2$
   & -- \\
 9 & $S_1^2-S_1S_3$
   & $b_3$
   & -- \\
\end{tabular}
\caption{List of coefficients used in the definition of the pion mass, the pion decay constant, and the scalar mass. It should be noted that not all coefficients are independent. The coefficients $a_{M,F}$ and $b_{M,F}$ are the standard results from chiral perturbation theory and they can be found in Table~\ref{table:gamma}. In the definition of $c_i$ the constant $n_\pi$ denotes the number of pions (the number of broken generators).}
\label{table:abc}
\end{table}
\begin{table}
\begin{tabular}{c|c|c}
 $\qquad$
 & $\quad SU(2) \times SU(2) \to SU(2)\quad$
 & $\quad\qquad SU(4) \to Sp(4)\qquad\quad$ \\
 \hline
 $\Gamma^K_1$ & $-\tfrac{1}{8}S_1^2+\tfrac{3}{2}S_2$
              & $-\tfrac{1}{8}S_1^2+\tfrac{3}{2}S_2$ \\
 $\Gamma^K_2$ & $2S_2-\tfrac{1}{2}S_4$
              & $2S_2-\tfrac{1}{2}S_4$ \\
 $\Gamma_4$   & $-\tfrac{1}{32}S_1^2+\tfrac{1}{8}S_1S_3-\tfrac{1}{2}\Gamma_5+\tfrac{1}{4}$
              & $-\tfrac{1}{64}S_1^2+\tfrac{1}{16}S_1S_3-\tfrac{1}{4}\Gamma_5+\tfrac{1}{8}$\\
 $\Gamma_5$   & $\tfrac{1}{4}$
              & $\tfrac{1}{4}$ \\
 $\Gamma_6$   & $\tfrac{1}{64}S_1^2+\tfrac{1}{16}S_3^2-\tfrac{1}{2}\Gamma_8+\tfrac{3}{32}$
              & $\tfrac{1}{128}S_1^2+\tfrac{1}{32}S_3^2-\tfrac{1}{4}\Gamma_8+\tfrac{5}{128}$ \\
 $\Gamma_8$   & 0
              & 0 \\
 \hline
 $a_M$        & $\tfrac{1}{2}$
              & $\tfrac{3}{4}$ \\
 $b_M$        & $16(2L_6^r+L_8^r)-8(2L_4^r+L_5^r)$
              & $16(4L_6^r+L_8^r)-8(4L_4^r+L_5^r)$ \\
 $a_F$        & $-1$
              & $-1$ \\
 $b_F$        & $8L_4^r+4L_5^r$
              & $16L_4^r+4L_5^r$ \\
\end{tabular}
\caption{The coefficients $\Gamma_i$ and $\Gamma_i^K$ used in Eq.~\eqref{eq:Lr} and Eq.~\eqref{eq:Kr} to cancel the one-loop divergences in the pion mass and decay constant. The two coefficients $\Gamma_5$ and $\Gamma_8$ are unconstrained for these two symmetry breaking patterns, and the values are simply chosen to coincide with \cite{Bijnens:2009qm}. The constants $a_{M,F}$ and $b_{M,F}$ encode the standard results from chiral perturbation theory and, in the general case, they can also be found in \cite{Bijnens:2009qm}.}
\label{table:gamma}
\end{table}

\subsection{Consistency checks}
We now consider several limits and checks to verify our results and to show that we recover known results from current-algebra. For example, when setting $S_i=0$ and $K_i^r=0$ we recover the known ChPT results. This follows from the algebraic structure of $\hat{m}_\pi^2$ and $\hat{f}_\pi$ together with the values listed in Table~\ref{table:abc} and Table~\ref{table:gamma}.

When the bare pion mass vanishes, i.e. $m_\pi^2\to0$, we also recover the chiral limit of the theory that requires $\hat{m}_\pi^2 = 0$ together with a finite value for $\hat{f}_\pi$. The vanishing of the renormalised pion mass arises from the fact that $J_{\pi\sigma\pi}$ satisfies
\begin{equation}
 J_{\pi\sigma\pi} = L_\sigma = \frac{1}{16\pi^2}\log\left(\frac{m_\sigma^2}{\mu^2}\right),
\end{equation}
in the chiral limit, together with the relation $a_5=-a_4$. The pion decay constant, in the chiral limit, does not vanish 
\begin{equation}
 \hat{f}_\pi = f_\pi + \frac{m_\sigma^2}{f_\pi}\left(b_4 + (b_5+b_6)L_\sigma - \frac{b_8}{32\pi^2}\right),
 \label{eq:fhatchiral}
\end{equation}
where in this limit we used 
\begin{equation}
 H_{\pi\sigma\pi} = -\frac{1}{16\pi^2}\frac{1}{2m_\sigma^2} \ .
\end{equation}
The result in Eq. \eqref{eq:fhatchiral} clearly shows that $\hat{f}_\pi$ and $f_\pi$ no longer coincide in the chiral limit, because of the corrections from the scalar field.

As mentioned in Section \ref{sec:lagrangian}, the renormalized coefficients $L_i^r$ and $K_i^r$ depend on the renormalization scale $\mu$ and this dependence can be used to perform another consistency check. The check consists in changing the scale from $\mu$ to $\tilde{\mu}$ in our results and show that this translates into  shifting the renormalized coefficients in the following way.
\begin{align}
 L_i^r(\mu) \to L_i^r(\tilde{\mu}) + \frac{\Gamma_i}{32\pi^2}\log\left(\frac{\tilde{\mu}^2}{\mu^2}\right) \ ,  \\
 K_i^r(\mu) \to K_i^r(\tilde{\mu}) + \frac{\Gamma_i^K}{32\pi^2}\log\left(\frac{\tilde{\mu}^2}{\mu^2}\right) \ . 
\end{align}
These relations hold in our results and rely on the specific combinations of $L_x$ and $J_{xyz}$ because both depend on the renormalization scale. Given all the above we are confident that our results are solid.

\subsection{Scalar mass and width}
\begin{figure}
 \begin{center}
  \includegraphics[width=0.2\linewidth]{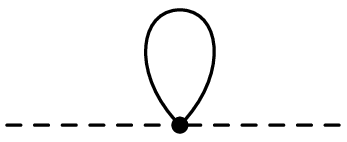}
  \hspace{2mm}
  \includegraphics[width=0.2\linewidth]{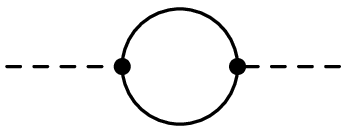}
  \hspace{7mm}
  \includegraphics[width=0.2\linewidth]{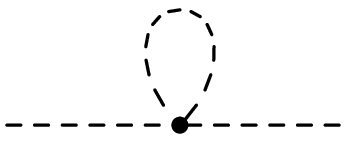}
  \hspace{2mm}
  \includegraphics[width=0.2\linewidth]{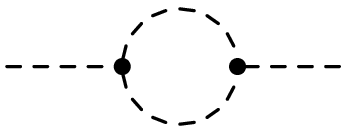}
  \caption{Loop diagrams contributing (at next-to-leading order) to the scalar mass. The solid lines are pions and the dashed lines are scalars.}
  \label{fig:scalar_diagrams}
\end{center}
\end{figure}

The next-to-leading order corrections to the two-point function for the scalar field stems from its coupling to the pions (the first two diagrams in Fig.~\ref{fig:scalar_diagrams}) and its self-interactions (the last two diagrams in Fig.~\ref{fig:scalar_diagrams}) coming from the $\sigma^4$ and $\sigma^3$  terms in the Lagrangian. Here we again define the physical scalar mass $\hat{m}_\sigma^2$ as the pole in the propagator. Using the previously introduced notation, the renormalised scalar mass reads
\begin{align}
\begin{split}
 \hat{m}_\sigma^2
 = m_\sigma^2
 + \frac{m_\sigma^4}{f_\pi^2}(c_1L_\sigma + c_2J_{\pi\pi\sigma} + c_3J_{\sigma\sigma\sigma} - 2K_3^r)
 &+ \frac{m_\pi^4}{f_\pi^2}(c_4L_\pi + c_5J_{\pi\pi\sigma}-2K_5^r) \\
 &\quad+ \frac{m_\pi^2m_\sigma^2}{f_\pi^2}(c_6L_\pi + c_7J_{\pi\pi\sigma}-2K_4^r),
\end{split}
\end{align}
with the coefficients $c_i$ listed in Table~\ref{table:abc}. To renormalise the scalar mass we choose the values of $\Gamma_i^K$ listed below and stress that they only depend on the number of pions $n_\pi$ but no other specific detail of the given breaking pattern.
\begin{align}
\begin{split}
 \Gamma^K_3 &= c_1+c_2+c_3  \ ,\\
 \Gamma^K_4 &= c_6+c_7 \ ,\\
 \Gamma^K_5 &= c_4+c_5  \ .
 \end{split}
	\end{align}
We notice that the scalar mass $\hat{m}_\sigma^2$ develops a branch-cut  at $m_\sigma=2m_\pi$ because the two pions in the second diagram of Fig.~\ref{fig:scalar_diagrams} can go on-shell. Above this value, the decay width of the scalar can be read off from the imaginary part of the mass.
\begin{equation}
 \Gamma
 = -\frac{c_2m_\sigma^4+c_5m_\pi^4+c_7m_\sigma^2m_\pi^2}{16\pi m_\sigma f_\pi^2}\sqrt{1-\frac{4m_\pi^2}{m_\sigma^2}} \ .
 \label{Width}
\end{equation}
In our approach we have two unknown LECs in the expression for the decay width. The first coefficient $S_1$ is the coupling between the scalar field and the kinetic term for the pions, while the second coefficient $S_3$ comes from the coupling between the scalar field and the pion mass term operator. For the $SU(2)_L \times SU(2)_R \to SU(2)_V$ case, the result for $\Gamma$ agrees with the expression obtained in \cite{Soto:2011ap} when using the identification $S_1=4c_{1d}$ and $S_3=0$.

\section{The many natures of the scalar\label{sec:natures}}
Because the scalar corrections have been introduced in a generic way, we can consider different origins for the scalar field, within the limits of the counting scheme. In practice each given nature corresponds to imposing relations among the $S_i$ couplings parametrising the scalar interactions.

\subsection{Dilaton}
An interesting class of theories is the one in which a light scalar emerges as  pseudo-dilaton   \cite{Sannino:1999qe,Dietrich:2005jn,Dietrich:2006cm,Goldberger:2008zz,Matsuzaki:2013eva,Golterman:2016lsd,Crewther:2012wd,Crewther:2013vea} for which the Lagrangian reads: 
\begin{equation} 
 \cL_2 = \frac{f_\pi^2}{4}\left[\tr{u_\mu u^\mu}\exp\left(\frac{2\sigma}{f_\pi}\right) + \tr{\chi_+}\exp\left(\frac{y\sigma}{f_\pi}\right)\right].
\end{equation}
Although we chose to use $f_\pi$ as the compensating scale for the pseudo-dilaton in the exponential, de facto, depending on the microscopic realization it can differ, and our results still apply. Expanding the exponential to second order we find that our $S_i$ are now related via 
\begin{equation}
 S_1 = S_2 = 2,\qquad S_3=y,\qquad S_4 = \frac{y^2}{2}.
\end{equation}
Here $y=3-\gamma^*$ with $\gamma^*$ being the anomalous dimension of the fermion mass in the underlying gauge theory. It is now evident that $\gamma^*$ is the only new parameter in the expression for the pion mass \eqref{eq:Mhat} and the pion decay constant \eqref{eq:Fhat} when the scalar field is a pseudo-dilaton.
\begin{align}
\begin{split}
 \hat{m}_\pi^2
 &= m_\pi^2
 + \frac{m_\pi^4}{f_\pi^2}(b_M+(a_M-2)L_\pi-(2-y)^2J_{\pi\sigma\pi})
 + \frac{m_\sigma^4}{f_\pi^2}(L_\sigma - J_{\pi\sigma\pi}) \\
 &\qquad+ \frac{m_\pi^2m_\sigma^2}{f_\pi^2}(a_6 - L_\pi - \tfrac{1}{2}(y+6)(y-2)L_\sigma+(4-2y)J_{\pi\sigma\pi})
 \label{dil-m}
\end{split}
\end{align}
\begin{align}
\begin{split}
 \hat{f}_\pi
 &= f_\pi + \frac{m_\pi^2}{f_\pi}(b_F + (a_F-\tfrac{3}{2})L_\pi + (2-y)J_{\pi\sigma\pi})
 + \frac{m_\sigma^2}{f_\pi}(b_4 + \tfrac{7}{2}L_\sigma - J_{\pi\sigma\pi}) \\
 &\qquad- \frac{H_{\pi\sigma\pi}}{f_\pi}(\tfrac{1}{2}(2-y)^2m_\pi^4 + \tfrac{1}{2}m_\sigma^4 - (2-y)m_\pi^2m_\sigma^2).
  \label{dil-f}
\end{split}
\end{align}
Here $a_6$ and $b_4$ contain the unconstrained coeffiecients $K_i^r$. In certain near-conformal theories, perturbation theory predicts that $\gamma^*\approx1$ is possible \cite{Ryttov:2016hdp}. In the limiting case $\gamma^*=1$ (or equivalently $y=2$) the expressions for $\hat{m}_\pi^2$ and $\hat{f}_\pi$ simplifies considerably, because the coefficient in front of several terms vanishes. We stress that the $\sigma$ field in this formulation is the fluctuation around the expectation value. For a discussion about how the expectation value depends on the low energy parameters for the dilaton and the pions we refer to \cite{Kasai:2016ifi}.

\subsection{Large-N Limit\label{subsec:largeN}}
It is well known that the pion decay constant squared is proportional to $N$ when the underlying dynamics, yielding the low energy effective theories, arises from an $SU(N)$ gauge theory with fermions transforming according to the fundamental representation of the theory. This means that in the large-$N$ limit the pion interaction strength vanishes. This feature is common to any meson that is predominantly made by a fermion-antifermion pair. Furthermore their mass is leading in $N$. Assuming that also the scalar singlet is a leading $N$ meson one finds
\begin{equation}
 f_\pi^2 \sim \cO(N),\qquad
 m_\sigma^2 \sim \cO(1).
\end{equation}
This counting is automatically encoded in our effective theory since, order-by-order, corrections are suppressed by factors of $f_\pi$.  It is possible to generalize the present formalism to encode different large-$N$ counting schemes arising when choosing, for example, fermions in different representations of the underlying gauge group as shown in \cite{Sannino:2015yxa}.

However in the strict large-$N$ limit one has also to take into account, for fermions in the fundamental representation, the fact that one more  state becomes parametrically light with $N$ i.e. the pseudo-scalar associated to the $U(1)$ axial anomaly. For a review on how to incorporate this state, and generalizations to different representations see \cite{DiVecchia:2013swa}.
 
\subsection{Linear Sigma Model}
The Lagrangian presented in Section \ref{sec:lagrangian} can be compared to the linear sigma model by properly matching the low-energy constants. To this end we consider the linear sigma model with $N$ pions $\phi^a$ and a single scalar field $\sigma$. With the notation $\Phi=(\sigma,\vec{\phi})^T$ we can write the Lagrangian of the linear sigma model manifestly invariant under a global $O(N+1)$ symmetry.
\begin{equation}
 \cL = \frac{1}{2}(\partial_\mu\Phi)^T(\partial^\mu\Phi)
     + \frac{1}{2}\mu^2(\Phi^T\Phi) - \frac{1}{4}\lambda(\Phi^T\Phi)^2,
\end{equation}
After the $\sigma$ field acquires a vacuum expectation value $v^2=\mu^2/\lambda$ the Lagrangian can be written as
\begin{equation}
 \cL = \frac{1}{2}(\partial_\mu\sigma)^2+\frac{1}{2}(\partial_\mu\vec{\phi})^2
     - \mu^2\sigma^2
     - \frac{1}{2}\lambda\sigma^2\vec{\phi}^2
     - v\lambda(\sigma\vec{\phi}^2+\sigma^3)
     - \frac{1}{4}\lambda(\sigma^4+\vec{\phi}^4).
\end{equation}
We observe that the pions are massless and it is understood that $\sigma$ and $\vec{\phi}$ are the fluctuations around the vacuum. After spontaneous symmetry breaking, the global symmetry of the Lagrangian has furthermore been reduced to $O(N)$. We can exploit the following homomorphisms to rewrite the linear sigma model in such a way that we have the same global symmetries as we do in the chiral Lagrangian.
\begin{align*}
 O(4) \to O(3) &\cong SU(2)_R \times SU(2)_L \to SU(2)_V \\
 O(6) \to O(5) &\cong SU(4) \to Sp(4).
\end{align*}
The linear sigma model can be written in terms of a matrix $\Sigma$ such that the above symmetries are manifest in the Lagrangian.
\begin{equation}
 \cL = \frac{1}{2D_R}\tr{\partial_\mu\Sigma^\dagger\partial^\mu\Sigma}
     + \frac{1}{2D_R}\mu^2\tr{\Sigma^\dagger\Sigma}
     - \frac{1}{4D_R^2}\lambda\tr{\Sigma^\dagger\Sigma}^2.
 \label{eq:ls_sigma}
\end{equation}
We define the  $\Sigma$ matrix as
\begin{equation}
 \Sigma = [(\sigma+v) + iX^a\phi^a]V,
 \label{eq:sigma_mat}
\end{equation}
where $X^a$ are the broken generators taken to be $D_R\times D_R$ square matrices normalized such that $\tr{X^aX^b}=D_R\delta^{ab}$. The matrix $V$ encodes the vacuum alignment. For $SU(2)_L \times SU(2)_R \to SU(2)_V$ it is the identity matrix, while for $SU(4) \to Sp(4)$ case \cite{Kogut:2000ek}, it is given by 
\begin{equation}
 V = \begin{pmatrix}
 0 & 0 & -1 & 0 \\
 0 & 0 & 0 & -1 \\
 +1 & 0 & 0 & 0 \\
 0 & +1 & 0 & 0
 \end{pmatrix}.
\end{equation}
 We are now ready to make the connection to our chiral Lagrangian by rewriting the $\Sigma$ matrix as
\begin{equation}
 \Sigma = (\sigma+v)UV,
\end{equation}
where $U=\exp(iX^a\phi^a/v)$ is a unitary matrix. To leading order in $1/v$ this definition coincides with the original definition in Eq.~\eqref{eq:sigma_mat}. The Lagrangian in Eq.~\eqref{eq:ls_sigma} can now be written as
\begin{equation}
 \cL = \frac{v^2}{2D_R}\left(1 + \frac{2\sigma}{v} + \frac{\sigma^2}{v^2}\right)\tr{\partial_\mu U^\dagger \partial^\mu U} + \frac{1}{2}(\partial_\mu\sigma)^2 - \lambda v^2\sigma^2\left(1 + \frac{\sigma}{v} + \frac{\sigma^2}{4v^2}\right).
 \label{eq:ls_final}
\end{equation}
By expanding the kinetic term for the pions we can now match this Lagrangian to our chiral Lagrangian in Eq.~\eqref{expandedL} and the scalar Lagrangian in Eq.~\eqref{scalarL} via the following identifications.
\begin{equation}
 f_\pi = v,\qquad
 S_1 = 2,\qquad
 S_2 = 1,\qquad
 m_\sigma^2 = 2\lambda v^2,\qquad
 S_5 = 1,\qquad
 S_6 = \frac{1}{4}.
\end{equation}
In Eq.~\eqref{eq:ls_final} the pions are massless and for this reason we cannot match the two coefficients $S_3$ and $S_4$. This could eventually be done by introducing an explicit breaking term in Eq.~\eqref{eq:ls_sigma}.

For completeness we notice that, at the classical level, one can relate the linear sigma model to the dilaton via the field redefinition
\begin{equation}
 \sigma \to f_\pi \left[\exp\left(\frac{\sigma}{f_\pi}\right) - 1\right].
\end{equation}

\subsection{Goldstone Boson}
If the scalar field is a Goldstone boson, then the effective theory is invariant under, at least, a shift symmetry $\sigma\to\sigma+a$. This implies that only derivative couplings are allowed in the Lagrangian. In our setup this corresponds to choosing all $S_i=0$ in which case we recover standard chiral perturbation theory results for the quantities computed here. However, scalar effects will appear at higher orders in the chiral expansion. One can allow for a controllably small breaking of the shift symmetry by requiring
\begin{equation}
 S_i\ll\cO(1).
\end{equation}
This will significantly reduce the effects from the scalar loops.

\subsection{QCD}
The present framework is directly applicable to QCD where different approaches point to the existence of a scalar state \cite{Sannino:1995ik,Harada:1995dc,Ishida:1995xx,Harada:1996wr,Harada:2003em,Black:2000qq,Oller:1998hw,Pelaez:2003dy,Pelaez:2006nj,Sannino:2007yp,Pelaez:2015qba,Caprini:2005zr,Hooft:2008we,GarciaMartin:2011jx,Parganlija:2012fy,Pelaez:2015zoa,Ghalenovi:2015mha}  with mass  $m_\sigma = 457$ MeV and width $\Gamma=558$ MeV, where the values are taken from reference \cite{Olive:2016xmw}.  

Several earlier and modern interpretations of the underlying nature of this state have been put forward in the literature \cite{Jaffe:1976ih,vanBeveren:1986ea,Janssen:1994wn,Tornqvist:1995kr,Tornqvist:1995ay,Kaminski:1993zb,Bolokhov:1993am,Morgan:1993td,Andrianov:1993dq,Svec:1995xr,Ishida:1997wn,Sannino:1995ik,Harada:1995dc,Ishida:1995xx,Harada:1996wr,Harada:2003em,Black:2000qq,Oller:1998hw,Pelaez:2003dy,Pelaez:2006nj,Sannino:2007yp,Pelaez:2015qba,Hooft:2008we,Parganlija:2012fy,Pelaez:2015zoa,Ghalenovi:2015mha}. These investigations seem to converge toward the presence of a large four-quark component of this state.  

Given an assumed nature of this state one can, using the present framework,  test it against experimental and lattice results, when available. For example already from the limited knowledge of the width and mass we can derive the relation 
\begin{equation}
 S_1 = -0.227S_3 + 2.535 \ ,
 \label{eq:S1qcd}
\end{equation}
which is in agreement with the expectation given in Eq. \eqref{eq:LKSest}. The contribution from $S_3$ is naturally suppressed by the small pion mass.

Assuming that the lightest massive scalar behaves as a pseudo-dilaton \cite{Sannino:1999qe,Sannino:2003xe} we have the further relations $S_1 = S_2 = 2$, $S_3 = y$ and $S_4 = y^2/2$ that from the previous constraint permits to determine $y = 2.357$ and consequently a would be fermion mass anomalous dimension of $\gamma^{\ast} = 3 -y = 0.643$. One can further test the relation, and consequently the limit, via its impact on the pion mass  \eqref{dil-m}  and decay constant \eqref{dil-f} at the NLO as well as in processes such as pion-pion scattering. In the linear sigma model limit \cite{Tornqvist:1997nr,Tornqvist:1997xm} we have  $S_1 = 2$  and $S_2 =1$ leading to the same prediction for $S_3$ but the mass and decay constant renormalise differently than in the pseudo-dilatonic limit. 

As for the  relevant interpretation in terms of a four-quark state \cite{Sannino:1995ik,Harada:1995dc,Ishida:1995xx,Harada:1996wr,Harada:2003em,Black:2000qq,Oller:1998hw,Pelaez:2003dy,Pelaez:2006nj,Sannino:2007yp,Pelaez:2015qba,Caprini:2005zr,Hooft:2008we,GarciaMartin:2011jx,Parganlija:2012fy,Pelaez:2015zoa,Ghalenovi:2015mha} one can envision different underlying realisations that range from this state emerging prevalently as bound state of pions to having a more compact wave-function at the quark level. Each of these possibilities will lead to specific predictions for the LECs that can, in principle, be obtained within model computations. For example, the mass of the sigma in a four-quark interpretation increases with the number of colors, see Fig.~5 of \cite{Harada:2003em}, modifying the large N counting in section \ref{subsec:largeN}. It would therefore be very exciting, in the future, to investigate these limits within the present framework.

\section{Conclusion} 
We added an isosinglet scalar to the chiral Lagrangian and determined the radiative corrections for the pion mass and decay constant. We also determined the quantum corrections for the two-point scalar function and determined its physical mass and width. The analysis is performed for two breaking patterns of immediate relevance for phenomenology and lattice simulations. Our analysis extends previous results and it embraces different physical realisations for the isosinglet, such as the dilaton, the (pseudo) Goldstone boson, the $\sigma$ state in QCD, and the linear sigma model. The results presented here can also be used to extrapolate a potentially light isoscalar mass to the chiral limit in lattice simulations.

\acknowledgments
The CP$^3$-Origins centre is partially funded by the Danish National Research Foundation, grant number DNRF90. M.H. is funded by a Lundbeck Foundation Fellowship grant.

\appendix
\section{One-loop Integrals}
\label{sec:loop_integrals}
In this appendix we list the one-loop integrals needed in our calculations. For the diagrams considered here we need a total of eight different integrals.
\begin{align}
 I_1 &= i\mu^\epsilon\int\frac{d^4k}{(2\pi)^4}\frac{1}{k^2-m^2}
      = A_0(m^2) \\
 I_2 &= i\mu^\epsilon\int\frac{d^4k}{(2\pi)^4}\frac{k^2}{k^2-m^2}
      = m^2A_0(m^2) \\
 I_3 &= i\mu^\epsilon\int\frac{d^4k}{(2\pi)^4}\frac{1}{[k^2-m_1^2][(k+q)^2-m_2^2]}
      = B_0(m_1^2,m_2^2,q^2) \\
 I_4 &= i\mu^\epsilon\int\frac{d^4k}{(2\pi)^4}\frac{q_\mu k^\mu}{[k^2-m_1^2][(k+q)^2-m_2^2]}
      = B_1(m_1^2,m_2^2,q^2) \\
 I_5 &= i\mu^\epsilon\int\frac{d^4k}{(2\pi)^4}\frac{k^2}{[k^2-m_1^2][(k+q)^2-m_2^2]}
      =  A_0(m_2^2) + m_1^2B_0(m_1^2,m_2^2,q^2)\\
 I_6 &= i\mu^\epsilon\int\frac{d^4k}{(2\pi)^4}\frac{q_\mu q_\nu k^\mu k^\nu}{[k^2-m_1^2][(k+q)^2-m_2^2]}
      = q^2B_2(m_1^2,m_2^2,q^2) \\
 I_7 &= i\mu^\epsilon\int\frac{d^4k}{(2\pi)^4}\frac{q_\mu k^\mu k^2}{[k^2-m_1^2][(k+q)^2-m_2^2]}
      = [m_1^2B_1(m_1^2,m_2^2,q^2)-q^2A_0(m_2^2)] \\
 I_8 &= i\mu^\epsilon\int\frac{d^4k}{(2\pi)^4}\frac{k^4}{[k^2-m_1^2][(k+q)^2-m_2^2]}
      = (m_1^2 + m_2^2 + q^2)A_0(m_2^2) + m_1^4B_0(m_1^2,m_2^2,q^2)
\end{align}
The solutions to the integrals are written in terms of the four functions listed below.
\begin{align}
 A_0(m^2) &= \frac{m^2}{16\pi^2}\left[\log\left(\frac{m^2}{\mu^2}\right) - R\right] \\[3mm]
 B_0(m_1^2,m_2^2,q^2) &= \frac{1}{16\pi^2}\left[1-R+\Jbar(m_1^2,m_2^2,q^2)\right] \\[3mm]
 B_1(m_1^2,m_2^2,q^2) &= \frac{1}{2}\left[A_0(m_1^2) - A_0(m_2^2) + (m_2^2-m_1^2-q^2)B_0(m_1^2,m_2^2,q^2)\right] \\[3mm]
 B_2(m_1^2,m_2^2,q^2) &= \frac{1}{2}\left[A_0(m_2^2) + (m_2^2-m_1^2-q^2)B_1(m_1^2,m_2^2,q^2)\right]
\end{align}
The unitarity corrections are parametrized by the function
\begin{align}
 \Jbar(m_1^2,m_2^2,q^2)
 &= \int_0^1 dx\log\left(\frac{xm_2^2+(1-x)m_1^2-x(1-x)q^2+i\epsilon}{\mu^2}\right) \\
\begin{split}
 &= \frac{1}{m_1^2-m_2^2}\left[m_1^2\log\left(\frac{m_1^2}{\mu^2}\right)-m_2^2\log\left(\frac{m_2^2}{\mu^2}\right)\right] - 1\\&\qquad+ \int_0^1 dx\log\left(\frac{xm_2^2+(1-x)m_1^2-x(1-x)q^2+i\epsilon}{xm_2^2+(1-x)m_1^2}\right).
\end{split}
\end{align}
In our results we also need the derivative of this function with respect to $q^2$.
\begin{equation}
 \Hbar(m_1^2,m_2^2,q^2)
 = \frac{\partial}{\partial q^2}~\Jbar(m_1^2,m_2^2,q^2)
 = \int_0^1dx~\left(\frac{x(x-1)}{xm_2^2+(1-x)m_1^2+x(x-1)q^2}\right)
\end{equation}

\end{document}